\documentclass[11pt,paper]{article}
\usepackage{graphicx}
\usepackage{amstext,amsfonts,amsbsy,amssymb,bbm}
\usepackage{amsmath,amscd,amssymb,latexsym}
\usepackage{a4wide}
\usepackage{times}
\newcommand {\be}[1]{\begin{eqnarray} \mbox{$\label{#1}$}  }
\newcommand{\ee}{\end{eqnarray}}

\newcommand{\pref}[1]{(\ref{#1})}

\newcommand\ie {{\it i.e.}, }
\newcommand\eg {{\it e.g. }}

\newcommand{\nn}{\nonumber\\}
\newcommand{\noi}{\noindent}

\newcommand\half{\frac 1 2 }

\newcommand{\pd}{\partial}

\newcommand{\mean}[1]{\left \langle #1 \right \rangle}

\newcommand{\ket}[1]{|#1\rangle}
\newcommand{\bra}[1]{\langle #1 |}

\newcommand{\com}[2]{\left[ #1, #2 \right]}

\newcommand{\ga}{ {\alpha} }
\newcommand{\gb}{ {\beta} }

\newcommand{\gG}{\Gamma}
\newcommand{\gd}{ {\delta} }
\newcommand{\gD}{ {\Delta} }
\newcommand{\gf}{\phi}
\newcommand{\gw}{ {\omega} }

\newcommand{\gs}{ {\sigma} }

\newcommand{\gQ}{\Theta}

\renewcommand{\ge}{ {\epsilon} }
\newcommand{\gl}{ {\lambda} }
\newcommand{\gr}{\rho}

\newcommand{\gk}{\kappa}

\newcommand{\sgn}{{\rm sgn}}

\begin{document}

\title{Fractional charge (and statistics) in Luttinger liquids\footnote{This is a written version of a talk given at the Joint IFIN-HH,ICTP, IAEA Workshop on Trends in Nanoscience: Theory, Experiment, Technology, in Sibiu, Romania, August 23-30, 2009.}}
\author{Jon Magne Leinaas\\
Department of Physics, University of Oslo, NO-0316 Oslo, Norway}

\date{November 5, 2009}

\maketitle

\begin{abstract} 
Charge fractionalization is the phenomenon where quasi-particle excitations in a many-particle system appear with non-integer values relative to the fundamental charge unit.  Examples of such systems are known from field theoretical models and from physical realizations.  Recently there has been an interest in charge fractionalization in one-dimensional systems described by Luttinger liquid theory. These are gapless systems and that gives rise to the question whether non-integer charges should be regarded as sharp in the same meaning as in a gapped system. In this talk I first give an introduction to charge fractionalization as an effect in gapped systems and discuss next in what sense the charge fractionalization effect is found in gapless Luttinger liquids. The talk is based on a recent paper with Mats Horsdal and Thors Hans Hansson \cite{LeinaasHorsdal09}.

\end{abstract}

\section{Introduction}
There are certain physical variables that appear, at the fundamental level, always with values that are multiples of a basic unit. A particular example is the electric charge which is quantized in multiples of the electron charge $e$. This discretization of electric charge is generally believed to have a quantum mechanical origin, even if there at this point is no completely general argument for this to be the case. In spite of the discreteness of electric charge at the fundamental level,  {\em charge fractionalization} is a phenomenon that may occur  in certain, sufficiently large many-particle systems. The fractionalization is then based on a redefinition of the charge operator, so that contributions from the ground state (vacuum state) of the many-particle system is suppressed and therefore only local contributions  are included which measure deviations from their background values. Such a suppression of the background is often the most sensible way to define physical variables in a low-energy description of the system. Under certain conditions the elementary, particle-like excitations of the a many-particle system, usually referred to as quasiparticles, will then appear with charges that have non-integer values when measured in terms of the original charge unit.

There exist several examples of how charge fractionalization may occur, both in theoretical models and in physical realizations. In particular, two cases of charge fractionalization in condensed matter systems have caught much attention. One is the case of half-integer charges in certain one-dimensional crystals characterized by a Peierls instability. In a field theoretical description of this system, the fractional charges are associated with soliton excitations \cite{JackiwRebbi76,Su79,Goldstone81,Niemi86}.
The second case has to do with quasi-particle excitations in the two-dimensional electron gas of the (fractional) quantum Hall effect. In this case the excitations appear with fractional charges as well as fractional statistics, and the values they take are determined by the ground state, with different rational values for each plateau of  the Hall conductivity \cite{Laughlin83,Halperin84,Arovas84}.  In addition  there are more recent suggestions that charge farctionalization may occur in two-dimensional graphene-like structures \cite{Hou07,JackiwPi07}.

There are certain features that seem to link these cases of charge fractionalization. The fractionalization can thus be viewed as a vacuum polarization effect, where the background defined by the many-particle system ``dresses" the quasiparticle with a non-integral number of the original particles, and the fact that there is in these systems a new, fractional unit of charge, reflects a certain property of the ground state which in the field theoretic description takes the form of a topological condition imposed on the fields. Thus, the presence of fractional quantum numbers may seem, quite generally, to be related to the presence of  some kind of {\em topological order} in the system \cite{Wen90,OshikawaSenthil06}, and this connection is one of the reasons that there have been much focus on charge fractionalization effects in many-particle systems. It is of interest to note that the physical dimensionality of the system may be important for such topological effects to take place, and the cases discussed so far are mainly restricted to systems in one and two dimensions.

Another feature that is shared by these examples of charge fractionalization is that the charged excitations are separated from the ground state by an energy gap. The gap suppresses the low frequency contributions to the background fluctuations which therefore may be filtered out in the appropriated ``smoothened" definition of the low energy variables. This is the basis for the new fractional units of charge to appear as quantum mechanically sharp in the low energy description.

Recently there has been an increased interest in the question of charge fractionalization in one-dimensional systems described by Luttinger liquid theory \cite{safi97,FisherGlazman97,Pham00,Steinberg08,LeHur08,Berg09,LeinaasHorsdal09}. This theory is assumed to apply under quite general conditions to many-particle systems in one dimension, and to be relevant to quasi-one-dimensional electron systems in quantum wires and carbon nano-tubes. The interest has been focussed not only on the theoretical analysis of the fractionalization effect in such systems, but also on how to demonstrate the effect experimentally. In fact charge fractionalization has been reported to have been seen in an experiment with electrons in a quantum wire \cite{Steinberg08}. However, one should note that the actual experiment shows this only in an average sense, and not in the form of individual, sharply defined fractional charges. In addition to this experiment there has quite recently been suggested another way to demonstrate the effect, by use of edge currents in a quantum Hall sample \cite{Berg09}.

The possibility of charge fractionalization in Luttinger liquids raises some questions  which I have recently discussed in a paper with Mats Horsdal and Hans Hansson \cite{LeinaasHorsdal09}, and which gives the background  for the present talk. These questions arise since the fractionalization effect in Luttinger liquids does not fit into the general picture outlined above. Thus, the system is gapless, and this makes it less clear in what sense fractional charges may be quantum mechanically sharp. There also seems to be no topological constraint on the possible values that such charges may take. Instead various fractional values may appear depending on the initial conditions which give rise to the fractionalization. 

The intention here is to discuss some of these questions. In the first part I will discuss charge fractionalization in the more conventional sense, by way of three examples. In the second part I outline how fractional charges may appear in a Luttinger liquid and discuss in what sense these can be viewed as sharp fractional values rather than being fractional in the average sense. The sensitivity of the charge values to different initial conditions are demonstrated by way of two examples, and finally in the summary I discuss the significance this has for our understanding of the effect. In the talk I avoid many details and refer to \cite{LeinaasHorsdal09} for a more complete discussion.

\section{Examples of charge fractionalization}
To illustrate the charge fractionalization effect I will briefly discuss three examples, first  the classic example  provided by the theoretical model of Jackiw and Rebbi \cite{JackiwRebbi76}.\\

\noi
{\large \it The Jackiw-Rebbi soliton}\\
 This fractionally charged state appears in a one-dimensional field theory with a scalar field $\phi(x,t)$ and a two-component Dirac field $\psi(x,t)$. In a semi-classical description the $\phi$-field is treated as a classical field, while the Dirac field is treated as a quantum field in the background of  $\phi(x)$. The field $\phi(x)$ satisfies a non-linear field equation, and this equation has a soliton solution which interpolates between two degenerate energy minima. In the background of the soliton the Dirac field has a localized zero energy solution, and the Dirac vacuum in this background is therefore degenerate, with two solutions corresponding to whether the zero energy state is occupied or not.

The interesting point is that the effect of the Dirac field is to ``dress" the soliton so it appears with half-integer charge, either $+ e/2$ or $- e/2$, with $e$ as the fermion charge. The soliton of the full theory, which can be interpreted as a physical particle of the theory, then has properties inherited from both the original fields $\phi$ and $\psi$, and the charge of this particle is not identical to that of the fundamental Dirac fermion.

A specific realization of the model is provided by the following Lagrangian for the $\phi$-field,
\be{lag}
{\cal L}_\phi=\half\pd_\mu \phi \,\pd^\mu \phi - \half (\gl/\gk)^2(\gk^2-\phi^2)^2
\ee
here written in relativistic form, with $\gl$ and $\gk$ as positive constants. The classical ground states correspond to the two constant fields $\phi_\pm=\pm \gk$, and any finite energy field has to approach one of these values asymptotically for $x\to\pm\infty$. A stationary soliton localized at the origin $x=0$ is described by the following solution of the field equation
\be{sol}
\phi_s(x)=\gk\tanh \gl x
\ee
It satisfies the boundary conditions $\phi_s(x)\to \pm \gk$ for $x\to\pm\infty$. There is also a corresponding anti-soliton given by $\phi_{\bar s}(x)=-\gk\tanh \gl x$, and there are other solutions where the solitons are shifted along the $x$-axis and where they are moving with constant velocity.

\begin{figure}[h]
\begin{center}
\includegraphics[width=7cm]{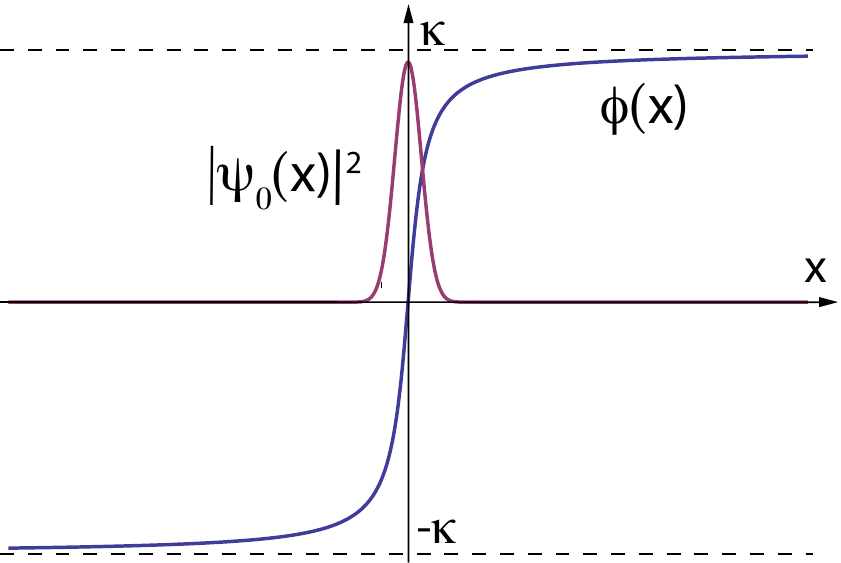}
\end{center}
\caption{\small The soliton solution of the Jackiw-Rebbi model. The blue curve shows the form of the scalar field $\phi_s(x)$ of the soliton, and the red curve shows the absolute square of the zero mode $\psi_0(x)$ of the Dirac field, which appears on the background of the scalar soliton field.\label{JackiwRebbi}}
\end{figure}

The Dirac field of the one-dimensional theory is described by a two component function $\psi(x)$, which is in the simplest case coupled linearly to the $\phi$-field. The Hamiltonian has then the form
\be{dirham}
H=\ga cp +\gb mc^2 \phi(x)/\gk
\ee
with $\ga$ and $\gb$ as anti-commuting 2x2 Dirac matrices, which can be represented by Pauli matrices, for example as $\ga=\gs_x$ and $\gb=\gs_y$. When the scalar field is in its ground state, either as $\gf=+\gk$ or $\gf=-\gk$, the Dirac Hamiltonian describes a free Dirac theory of fermions with mass $m$ and the vacuum state corresponds to a filled Dirac sea, where all negative energy states are occupied. However, in the background of the soliton the mass term tends to zero at the center of the soliton where the scalar field vanishes, and in the neighborhood of this point the localized zero energy state is found. It has the explicit form
\be{zero}
\psi(x)=N\exp(-\frac{mc}{\hbar\gk}\int_0^x dy \phi_s(y))\chi=N\exp[-\frac{mc}{\hbar\gl} \ln(\cosh \gl x)]\chi
\ee
with $N$ as a normalization factor and $\chi$ as a two-component spinor which satisfies $\ga\gb\chi=-i\chi$.
For the Dirac vacuum in the background of the soliton field $\phi_s$ this implies the two-fold degeneracy where all negative energy states are filled, but where the zero-energy state \pref{zero} may either be occupied or empty. The integrated charges of these two states can be evaluated and are found to be $\pm e/2$, and these are interpreted as the charges of the solitons dressed by the Dirac vacuum.

The charge values $\pm e/2$ of the solitons are determined by topological properties of the fields, and are therefore insensitive to local changes in the background field $\phi$. What is important is the asymptotic behavior of the field, with interpolation between the two minima. A further analysis of the soliton solutions shows that the half-integer charges are quantum mechanically sharp \cite{Kivelson82,Rajaraman82}.\\

\begin{figure}[h]
\begin{center}
\includegraphics[width=7cm]{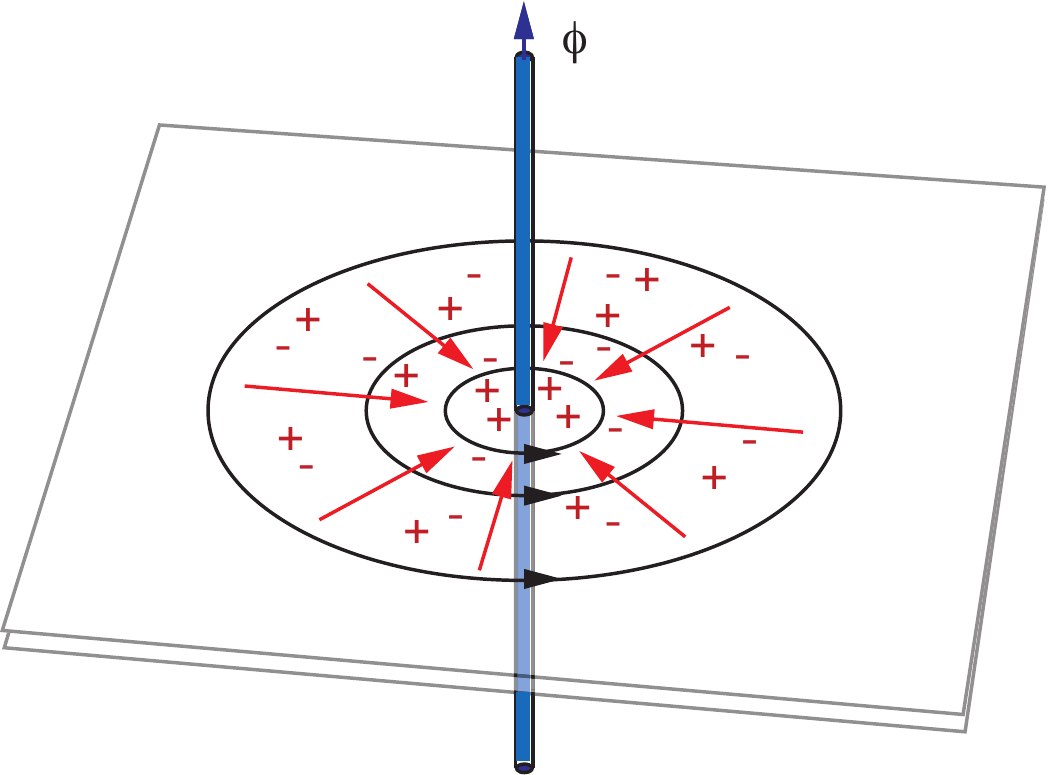}
\end{center}
\caption{\small The dressed flux tube in the 2-dimensional Dirac vacuum. The figure shows schematically the effect of the magnetic flux on the vacuum fluctuations, due to the presence of the Aharnov-Bohm field of the flux tube.  A surplus charge is attracted to the flux tube and a vacuum current is created that encircles the tube.  \label{String}}
\end{figure}

\noi
{\large \it The dressed flux tube}\\
In this case we consider the effect of a thin magnetic flux tube that penetrates a two-dimensional plane with Dirac fermions. The fermions are considered as spin polarized and are described by a two-component Dirac field. If the flux tube is assumed to be point-like in the plane, the stationary solutions of the Dirac equation can be found and charge and current densities of the vacuum state in the background of the flux tube can be determined \cite{Jaroszewicz86,Polychronakos87,Flekkoy91}. Also in this case there is a polarization of the Dirac vacuum due to the presence of the magnetic flux, and this dresses the flux with charge and creates a vacuum current that encircles the flux tube. These effects are schematically pictured in Fig.\ref{String}.

An interesting point is that the total charge attracted to the flux tube is  directly proportional to the magnetic flux $\gf$,
\be{fluxcharge}
q=-\half{\gf\over{\gf_0}} e
\ee
as long as $|\gf|<\gf_0$, with $\gf_0=h/e$ as the flux quantum and $e$ as the charge of the Dirac fermions. For larger values of $\gf$ the charge is a piecewise linear function of $\gf$, in a periodic fashion. This linear relation between $q$ and $\gf$ reflects again a topological property of the charge fractionalization. Thus, the integrated charge is determined by the boundary condition satisfied by the Dirac fields, which in turn depends only on the total magnetic flux inside the boundary. Therefore details of the local form of the magnetic flux are unimportant for the value of the integrated vacuum charge that is attracted to the flux tube \cite{Niemi86}.\\

\noi
{\large \it The Laughlin quasiparticle}\\
In the quantum Hall effect, where a two-dimensional electron gas is subject to a strong perpendicular magnetic field, the elementary charged excitations for a plateau state with fractional filling are known to carry a fraction of a unit charge and to be anyons rather than bosons or fermions. This was first discussed by R.B. Laughlin \cite{Laughlin83} and the fractional charge carried by the quasiparticles was later verified experimentally in tunneling experiments \cite{Goldman95,Saminadayar97,Picciotto97}.

\begin{figure}[h]
\begin{center}
\includegraphics[width=7cm]{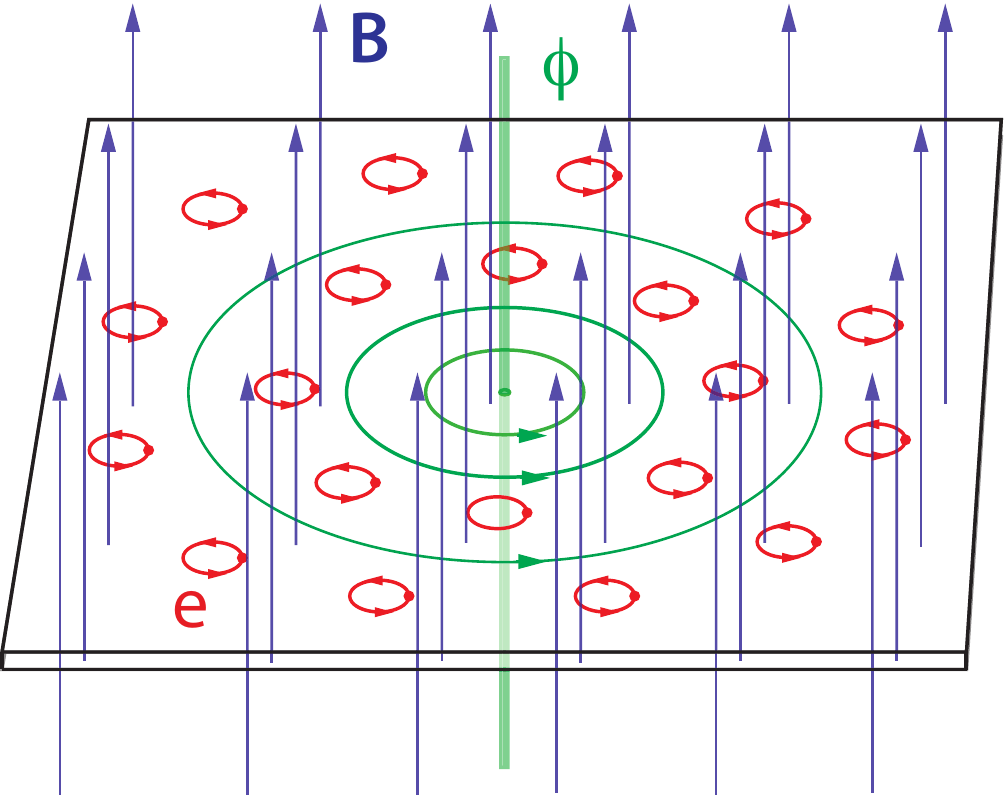}
\end{center}
\caption{\small Creation of a fractionally charged quasiparticle in the two-dimensional  electron gas of a quantum Hall system. A strong magnetic field $\bf B$ (illustrated by the blue arrows) penetrates the plane of the electrons and trap them in circular orbits in the lowest Landau level. The electrons form an incompressible fluid for certain fractional fillings. The figure illustrates the effect of piercing the fluid with a magnetic flux tube (green). When the flux $\phi$ is turned up a charged hole is created, and when a full flux quantum is reached a fractionally charged quasiparticle has been produced.\label{2DEG}}
\end{figure}

The simplest case of such a charge fractionalization effect appears in the Laughlin state at 1/3 filling of the lowest Landau level. A way to understand the appearance of the fractionally charged excitation is to imagine the two-dimensional electron gas pierced with a thin magnetic flux tube in a similar way as outlined in the previous example. The effect of this additional magnetic flux is to create a vortex in the electron gas which repels electron charge from the flux tube and creates a current encircling the tube \cite{Laughlin83}. In this case the charge and current created by the flux tube can be viewed as a polarization effect of the ``sea" of electrons in the ground state, in much the same way as the polarization of the Dirac vacuum in the previous example. Fig.\ref{2DEG}  shows a schematic picture of the situation.

When the magnetic flux is increased to be exactly one flux unit, then the non-local effect that the flux tube has on the electrons will disappear. It becomes in this sense invisible to the electrons and the ground state returns to the original ground state of the 1/3 state without the additional flux of the tube. However, if the flux is adiabatically turned up, it will leave a footprint behind.  A hole in the electron density is created even if the direct influence of the flux on the electrons disappears, and this hole represents a true quasiparticle excitation of the two-dimensional electron. The use of the imaginary flux tube is simply a theoretical construction to derive the form of the quasihole, but for the physical system the fractionally charged excitation can instead be created by changing for example the strength of the background magnetic field.

An interesting point of Laughlin's analysis is the suggestion of a simple many-electron trial wave function for the 1/3 ground state and for the quasi-hole excitation. With $z_i$ as complex coordinates for the electrons in the plane (measured in units of magnetic length), the quasi-hole wave function has the form
\be{qh}
\psi(z_1,z_2,...,z_N)={\cal N}\prod_i(z_i-z_0)\,\psi_0(z_1,z_2,...,z_N)
\ee
with $\cal N$ as a normalization factor, $z_0$ as the quasihole coordinate and $\psi_0(z_1,z_2,...,z_N)$ as the N-electron ground state,
\be{gs}
\psi_0(z_1,z_2,...,z_N)={\cal N'}\prod_{i<j}(z_i-z_j)^3\exp(-{1\over 4} \sum_k |z_k|^2)
\ee
 For this trial function the properties of the quasihole such as fractional charge and fractional statistics can readily be checked \cite{Arovas84}. Numerical studies have also demonstrated the quantum mechanical sharpness of these fractional charges \cite{Kjonsberg99}.
 
 As already pointed out there are several features these three examples of charge fractionalization outlined above have in common. The first one is that the local concentration of a fraction of a unit charge can be viewed as a polarization effect of the medium in which the excitation is created. In the first two examples it is a vacuum polarization effect whereas in the third example it is a polarization effect of the two-dimensional electron gas. In all three cases this polarization can be viewed as due to the presence of a background field, which is in the first case based on a semiclassical treatment and in the third example on the use of a fictitious magnetic flux. The second point is that the fractionalization effect in these cases may be viewed as having a topological origin. In the first case the topology is related to the twist in the scalar soliton field when it interpolates between the two minima, and this affects the positive and negative energy states of the Dirac field in an asymmetric way. In the two other cases the topological effect can be ascribed to the asymptotic behavior of the vector potential introduced by the magnetic field of the flux tube. The topology gives a robustness to the charge fractionalization effect, and makes the values of the fractional charges independent of local details \cite{Niemi86}.
 
 A third point to mention is that the fractional charges in the cases we have discussed are to be regarded as sharp in the quantum mechanical sense. This has been explicitly demonstrated by evaluating the variance of the local charge in a region that  includes the a fractionally charged excitations. The sharpness of the charge can be linked to the presence of a gap in the energy spectrum, where in the first two cases the gap is determined by the mass of the Dirac fermions, and in the last example it is given as the  energy gap from the ground state to the lowest charged excitation. This sharpness of the charge is important for the interpretation of the  fractionally charged  excitations as quasiparticles with modified particle properties relative to those of the fundamental particles of the theory.
 
The question of sharp fractional charges in Luttinger liquids is less clear, precisely because of the lack of a gap in the spectrum. The idea is to discuss this question, with reference to the results of the paper \cite{LeinaasHorsdal09}, but first I would like to discuss in general terms the meaning of notion of sharp fractional charge, when this is applied to excitations in a many-particle system.

\section{Fractional charge versus fractional probability}

For a single particle the question of sharpness of an observable can readily be answered. If the statistical variance of the observable vanishes, which means that the quantum state is an eigenstate of the observable, we refer to the observable as sharp, otherwise it is not. This can be illustrated by comparing two situations, where in the first one (case A) the wave function of a particle with charge $e$ is divided in two parts, with equal probability for the particle to reside inside one of the two boxes. In the other situation (case B) the charge itself is split in two, where half the charge, $e/2$, is located in each of the boxes.  

Let us consider a charge operator $Q_1$ which measures the charge only in box 1. In both situations the expectation value of the operator is $e/2$, but the variance is different. We have for the two cases
\be{AB}
{\text case\; A:} \;\;  (\gD Q_1)^2=e^2/4\,,\quad {\text case\; B:} \;\;  (\gD Q_1)^2=0
\ee
Clearly, in the first case it is the probability rather than the charge that is divided, so that the charge is either $0$ or $e$, both with probability $1/2$. This means that the quantum fluctuations of the charge are large and that is reflected in the non-vanishing value of the variance $(\gD Q_1)^2$. In the other case it is however the charge itself that is divided so that the charge in box 1 is $e/2$ with probability $1$. This gives the vanishing variance of the charge operator $Q_1$. Obviously charge fractionalization, if this is a meaningful concept, should correspond closer to {\em case B} than to {\em case A}.

However, in a many-particle system the distinction between the cases is not so obvious. The reason is that the charge to be measured sits on the top of the many-particle ground state, and even if the mean value of the charge density of the ground state vanishes there are in general large fluctuations. If thus a local charge operator is introduced which measures the charge within a well-defined region of space, this will for the many-particle states always have non-vanishing variance due to fluctuations in the charges along the boundary of the region. However, if the charge fluctuations are resolved in their frequency components, then the presence of an energy gap $\gD$ for the charged excitations will effectively suppress the low frequency components of the fluctuations. This implies that in a low-energy approximation to the full theory, the charge fluctuations may be filtered out.

\begin{figure}[h]
\begin{center}
\includegraphics[width=14cm]{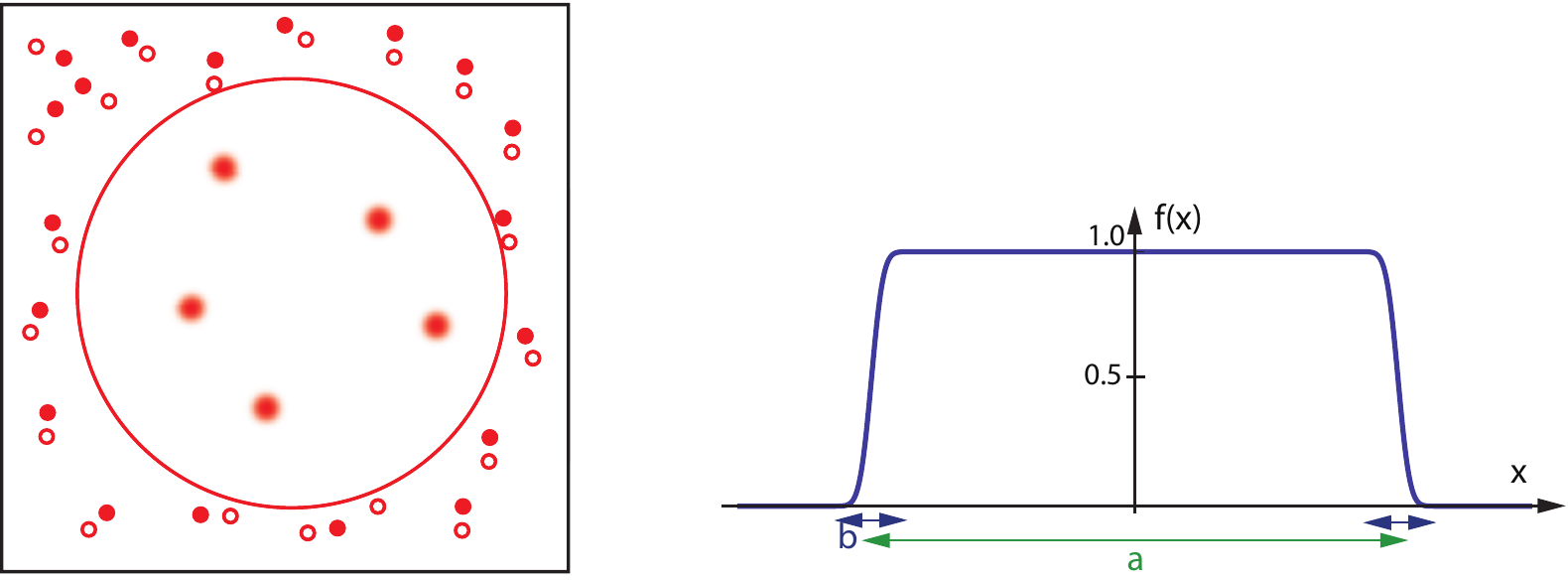}
\end{center}
\caption{\small Low frequency filter for the charge fluctuations. The charge fluctuations are to the left pictured as the presence of virtual particle-hole pairs in the ground state. The idea of a low frequency filter (inside the circle) is to suppress the fluctuations that are irrelevant for the low energy description, and thereby to identify the quasiparticles with sharply defined (fractional) charges. To the right the form of the particular sampling function with soft edges is shown, which is introduced in the text. \label{Filter2}}
\end{figure}

Let us restrict the discussion to the one-dimensional case, with the many-fermion system on the $x$-axis. A way to filter out the high frequency fluctuations is to define the local charge with a soft sampling profile for the charge. A specific way to do this is to introduce a Gaussian blur in the definition of the charge density\footnote{The space averaging actually suppresses the short wavelength rather than the high frequency contributions. In one dimension the difference is not important, but in higher dimensions an explicit frequency cutoff may be needed, as discussed in \cite{Goldhaber91}.}.
We write the averaged density as
\be{blur}
\tilde\gr(x)={1\over b\sqrt\pi} \int dx' e^{-(x-x')2/b^2}\gr(x')
\ee
with $b$ as a softening parameter.  The corresponding expression for the charge sampled in the interval $(-a/2,a/2)$ is
\be{err}
Q(a,b)=\int_{-a/2}^{a/2}\tilde\gr(x)=\int_{-\infty}^{\infty} dx f(x;a,b) \gr(x)
\ee
where the sampling function $f(x;a,b)$ is given by
\be{sampling}
f(x;a,b)=\half[{\rm erf}({{x+a/2}\over b})-{\rm erf}({{x-a/2}\over b})]
\ee
with erf$(x)$ as the error function. The sampling function is then essentially constant, with value $1$ over the interval $(-a/2,a/2)$, and drops exponentially fast to zero, with $b$ as a decay parameter, outside this interval, as shown in Fig.\ref{Filter2}.

For the charge fluctuations $\gD Q(a,b)$ the parameter  $b$ will effectively define a high frequency cutoff, while the energy gap $\gD$ acts as a low frequency cutoff, so that  if $1/b<<\gD/\hbar c$ there will be a suppression of the fluctuations in the whole frequency range. This can be demonstrated explicitly by considering charge fluctuations of the one-dimensional Dirac vacuum discussed earlier in the first example. In that case the energy gap of the system is defined by the fermion mass, $\gD=mc^2$. With the charge operator $Q(a,b)$ defined as above,  the variance $\gD Q(a,b)$ is given by the following integral over Fourier modes,
\be{varq}
\gD Q(a,b)^2={1\over{2\pi^2}}\int_{-\infty}^{\infty}dk \int_{-\infty}^{\infty}dk'
\left[1-\frac{(mc/\hbar)^2-kk'}{\sqrt{((mc/\hbar)^2+k^2)((mc/\hbar)^2+{k'}^2)}}\right]\frac{\sin^2(a(k+k')/2)}{(k+k')^2}
e^{-\half b^2(k+k')^2}\nn
={1\over{4\pi^2}}\int_{-\infty}^{\infty}dK \int_{-\infty}^{\infty}d\gk
\left[1-\frac{\gk^2-K^2+4(mc/\hbar)^2}{\sqrt{(\gk^2-K^2+4(mc/\hbar)^2)^2+8K^2(mc/\hbar)^2)}}\right]\frac{\sin^2(aK/2)}{K^2}
e^{-\half b^2K^2}\nn
\ee
where in the last expression the substitutions $K=k+k'$ and $\gk=k-k'$ have been done.

The exponential function effectively limits the $K$-integral so that $|K|\lesssim 1/b<< mc/\hbar$, and keeping only the leading contribution in an expansion of the integrand in powers of $K\hbar/mc$, we get for the integral
\be{varq2}
\gD Q(a,b)^2&\approx&{{(mc/\hbar)^2}\over{\pi^2}}\int_{-\infty}^{\infty}{{d\gk}\over{(\gk^2+4(mc/\hbar)^2)^2}}\int_{-\infty}^{\infty}dK 
\sin^2(aK/2)
e^{-\half b^2K^2}\nn
&=&\frac{\sqrt{\pi}\hbar}{32mc\,b}(1-e^{-a^2/4b^2})
\ee
This expression shows that the variance of the local charge can be made negligibly small by choosing $b$ sufficiently large, with $mc\,b/\hbar >> 1$.

Also the charge fluctuations of the soliton can be suppressed by the same smooth sampling function and in this sense the half-integral charge of the soliton is to be regarded as quantum mechanically sharp \cite{Kivelson82,Rajaraman82}.

It is clear from this result that the suppression of the charge fluctuation by the use of a smooth sampling function depends on the presence of a mass gap. Also this we may demonstrate explicitly, now by taking the limit $m\to 0$. In this limit the integrals in the definition of  $\gD Q(a,b)^2$ can be solved exactly 
\be{varq3}
\gD Q(a,b)^2&=&{{2}\over{\pi^2}}\int_{-\infty}^{\infty}dK 
{{\sin^2(aK/2)}\over K}
e^{-\half b^2K^2}\nn
&=&{1\over{2\pi^2}}({a\over b})^2\;  _2F_2(1,1;{3\over 2},2;-\half({a\over b})^2)
\ee
with $_2F_2$ as a generalized hypergeometric function. For a large ratio $a/b$ the variance shows a logarithmic behavior,
\be{varq4}
\gD Q(a,b)^2\approx{1\over{\pi^2}}\ln({a\over b})
\ee
The divergent behavior of the variance for $b\to 0$ is present also in the case $m\neq 0$, as the limit where  the edges of the sampling function becomes sharp. However, with $m=0$, it is the size $a$ of the sampling region rather than the mass $m$ that defines the low frequency cutoff . Thus, in the massless case there is an infrared divergence when $a\to\infty$ which is not cured by the presence of the smoothness parameter $b$.

This effect is relevant for the Luttinger liquid theory, due to the lack of an energy gap. Also there the ground state fluctuations cannot simply be suppressed by a smooth edge, and the discussion of whether local charges can be quantum mechanically sharp therefore depends on viewing the question of sharpness in a different way. 

\section{The one-dimensional Luttinger liquid}

\begin{figure}[h]
\begin{center}
\includegraphics[width=7cm]{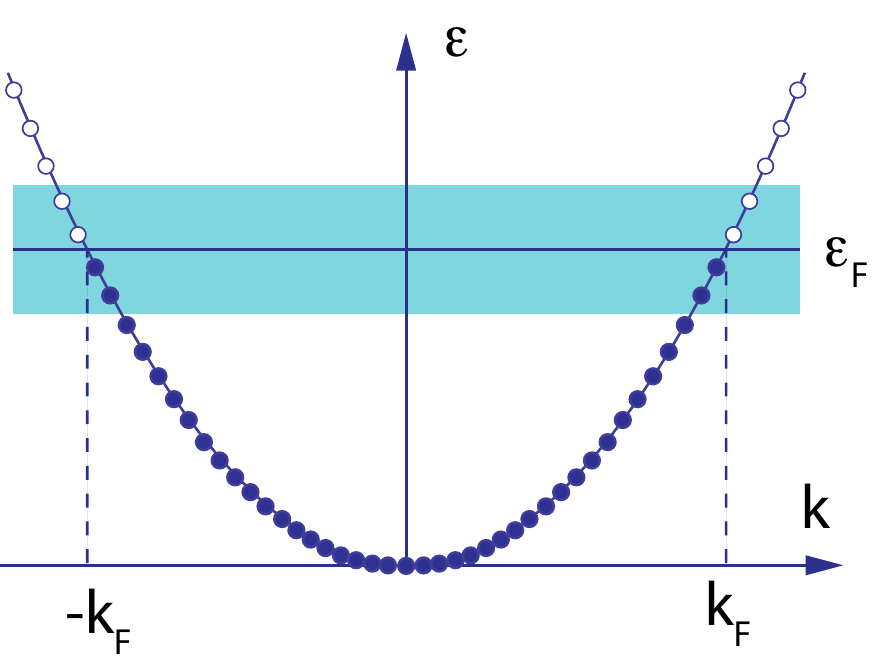}
\end{center}
\caption{\small The low-energy band of the one-dimensional fermion system. The figure shows the quadratic dispersion of the non-interacting system with all single-particle states occupied between the two Fermi points with momenta $\pm k_F$ (filled blue dots) and all states outside these points unoccupied (open blue dots). In the low energy approximation excitations are restricted to energies close to the Fermi energy $\ge_F$ (light blue band). For the interacting system the picture is essentially the same, but then with interactions between the fermions which are either close to the same Fermi point or close to opposite points.   \label{Harmonic}}
\end{figure}

Let me now focus on a many-fermion system which is described by Luttinger liquid theory in the low energy approximation. The system is non-relativistic, and is assumed to be clean in the sense that no background potential affects the particles, except the one that confines them to one dimension. The fermions are considered as spinless, and if the interaction between the particles is neglected the ground state corresponds to the situation illustrated in Fig.\ref{Harmonic} where all single particle states with energies below the Fermi energy $\ge_F$ are occupied and all the states above $\ge_F$ are unoccupied. 

The interacting system will under quite general conditions have a ground state which is qualitatively like the that of the non-interacting system. Thus the Fermi sea, defined by the ground state, is only modified in the neighborhood of the Fermi surface, which in the one-dimensional case is restricted to two points. A low-energy approximation to the interacting theory is furthermore based on the assumption that only excitations limited to a an energy interval around the two Fermi points which is small in the sense $\gD\ge<<\ge_F$ are of importance. This restriction is indicated by the light blue band in the Fig.\ref{Harmonic}, and a linearized approximation to the dispersion formula $\ge=\hbar^2k^2/2m$ is meaningful in this energy interval.

The linearized theory defines the Tomonaga-Luttinger model, which treats the fermions close to the two Fermi points as two different species \cite{Tomonaga50,Luttinger63}. When the system is confined to a circle of length $L$ the Hamiltonian is
\be{fermham}
H=v_F\hbar \sum_{\chi ,\,k}(\chi k-k_F):c_{\chi,\, k}^{\dag}\,c_{\chi,\, k}:+\frac{1}{2L} \sum_{\chi,\,q}(V_1(q)\rho_{\chi,\,q}\rho_{\chi,\,-q}+V_2(q)\rho_{\chi,\,q}\rho_{-\chi,\,-q})
\ee
with $\chi=\pm $ as a parameter which distinguishes the two species, $c_{\chi,\, k}$ and $c_{\chi,\, k}^\dag$ are fermion annihilation and creation operators and $\rho_{\chi,\,q}$ are Fourier components of the charge densities of the two species.  The parameter $v_F$ is the Fermi velocity and $k_F$ the Fermi momentum. The interaction is here separated in two parts, with $V_1$ as the interaction between pairs of particles close to the same Fermi point, and $V_2$ as the interaction between particles at opposite Fermi points. A {\em local} interaction between the fermions introduces the restriction $V_1=V_2$, however for more general (non-local) interactions the two interaction potentials may be different.  
Although the $k$ quantum number is in the low energy approximation restricted to small deviations from $\pm k_F$, this restriction is not essential since the low energy sector of the theory is not affected by an extension of the allowed values of $k$. This extended theory describes a system of massless Dirac particles.
 
One should note that the Hamiltonian \pref{fermham} has no matrix element that changes the relative number of fermions at the two Fermi points. This implies that there are two conserved fermion numbers $N_\chi$, that both take integer values. 

The one-dimensional fermion theory is different from the theory in higher dimensions in one important aspect. In one dimension the theory can be fully described in terms of bosonic variables, which are essentially the charge and current densities. The {\em interacting} fermion theory can, quite remarkably, be written as a {\em free} boson theory of the form
\be{hamdiag}
H=\sum_{q\neq 0}\hbar\gw_q \,b_q^{\dag}b_q
+ {{\pi\hbar}\over{2L}}(v_N N^2+v_JJ^2)
\ee
where the bosonic operators $b_q$ and $b_q^\dag$ are linear in the charge and current densities. The interactions are now hidden in the 
frequency $\gw_q$, given by
\be{freq}
\gw_q=\sqrt{\left(v_F+\frac{V_1(q)}{2\pi\hbar}\right)^2-\left(\frac{V_2(q)}{2\pi\hbar}\right)^2}\;|q|
\ee
and in the two velocity parameters
\be{velpar}
v_N=v_F+\frac{1}{2\pi\hbar}(V_1(0)+V_2(0))\,,\quad
v_J=v_F+\frac{1}{2\pi\hbar}(V_1(0)-V_2(0))
\ee
In the above expressions $N=N_++N_-$ is the total fermion number  and $J=N_+-N_-$ is the conserved current quantum number.

Under the assumption that the interactions are of sufficiently short range, the $q$ dependence of the Fourier transformed potentials can be neglected,
 $V_1(q)\approx V_1(0)$ and $V_2(q)\approx V_2(0)$. The dispersion of the bosonic field is then linear in $|q|$ and the Hamiltonian can be given the following field theoretic form   
\be{hfield}
H={u\over2}\pi\hbar \int_0^L dx \left[g^{-1}\left(\pd_x\Phi\right)^2+g\left(\pd_x\Theta\right)^2\right]
\ee
with $u=\sqrt{v_N v_J}$ and $g=\sqrt{v_J/v_N}$. 
In this formulation $\Theta$ and $\pd_x\Phi=\pd\Phi/\pd x$ (or alternatively $\Phi$ and $\pd_x\Theta$) are regarded as conjugate field variables, with the basic field commutator given as
\be{fieldcom}
\com{\Theta(x)}{\pd_x\Phi(x')}=\com{\Phi(x)}{\pd_x\Theta(x')}=\frac{i}{\pi}\gd(x-x')
\ee
The field $\pd_x\Phi$ can be identified with the fermion number density $\gr(x)$ of the original description  while $\pd_x \Theta$ is proportional to the current density $j(x)$.

The fields $\Theta(x)$ and $\Phi(x)$ are related to the bosonic creation and annihilation operators in the following way
\be{fieldop}
\Theta(x)&=&\Theta_0+{x\over L}J-i\sum_{q\neq 0}{1\over\sqrt{2\pi Lg|q|}}(b_q e^{iqx}-b_q^\dag e^{-iqx})\nn
\Phi(x)&=& \Phi_0-{x\over L}N+i\sum_{q\neq 0}\sqrt{{g\over2\pi L|q|}}\,\sgn(q)(b_q e^{iqx}-b_q^\dag e^{-iqx})
\ee
with $\Theta_0$ and $\Phi_0$ are $x$-independent operators that generate changes in the fermion numbers $N$ and $J$.

There is one important difference between the two representations of the Hamiltonian given by \pref{hamdiag} and \pref{hfield}. In \pref{hamdiag} there is an explicit separation of the $q=0$ contribution to the Hamiltonian, which depends on the conserved fermion numbers $N$ and $J$, and the $q\neq 0$ contribution which involves the bosonic variables $b_q$ and $b_q^\dag$. In \pref{hfield}, on the other hand, the $N$ and $J$ dependent part of the Hamiltonian  does not appear explicitly, but is hidden in the {\em zero mode}, which is the non-propagating $q=0$ part of the theory. This mode is linked to topological properties of the fields, which are reflected in the quasi-periodic condition 
\be{qper}
\Theta(x+L)=\Theta(x)+J\,,\quad \Phi(x+L) = \Phi(x)-N
\ee
In the field theoretic description the fermions can therefore be interpreted as topological excitations of the $(\Phi,\Theta)$ fields, and this gives an interesting, unified description of both the fermionic and bosonic variables. However, one of the points to be stressed is that for the discussion of sharpness of the fractional charges a distinction between the $q=0$ and the $q\neq 0$ contributions needed \cite{LeinaasHorsdal09}.

One should note that in the linearized description \pref{hfield} the two interactions $V_1$ and $V_2$ give rise to only one dimensionless interaction parameter $g$, while the remaining effect of the interactions is to rescale the velocity of propagation $u$. The case of non-interacting fermions, with $V_1=V_2=0$, gives $g=1$, and in the following I will refer to this value of $g$ as describing the {\em non-interacting} system, even if in reality it requires only $V_2=0$. Also note that in the following I will refer to {\em charge} as equivalent to {\em fermion number} without including the charge unit $e$ and I will measure fermion numbers relative to those of a fixed ground state, which then is assigned the values $N_\pm=0$.

\section{Chiral separation and fractional charges}
The fields $\Theta(x)$ and $\Phi(x)$ satisfy a one-dimensional wave equation, and can be separated in a natural way in two parts, its right- and left-moving components. These components, which can be defined as
\be{chicomp}
\Theta_\pm(x)= \Theta(x)\mp{1\over g}\Phi(x)
\ee
satisfy the linear differential equations
\be{lin}
(\pd_t\pm u \pd_x)\Theta_\pm(x)=0
\ee
so that $\gQ_+(x)$ describes the right-moving and  $\gQ_-(x)$ the left-moving modes. These two types of solutions of the field equation are also referred to as the positive and negative chirality modes.

The Fourier expansion of the chiral fields have the following form
\be{chiralfield}
\Theta_\pm(x)=\Theta_{0\pm}\pm{2x\over g L}Q_\pm-i\sum_{\pm q> 0}{\sqrt{2\over \pi Lg|q|}}(b_q e^{iqx}-b_q^\dag e^{-iqx})
\ee
with the zero mode operators
\be{zerofields}
\Theta_{0\pm}=\Theta_0\mp{1\over g}\Phi_0 \,,\quad
Q_\pm = \half(N\pm g J)
\ee
These operators satisfy the $g$-independent commutation relations
\be{comut}
\com{\Theta_{0i}}{Q_j}=-{i\over\pi} \gd_{ij} 
\ee
with $i,j=\pm$.

The operators $Q_\pm$ have a natural interpretation as {\em chiral} charge components, with $Q_+$ as the part of the total charge associated with the right-moving modes and $Q_-$ as the part associated with the left-moving ones. In the non-interacting case, with $g=1$, we have $Q_\pm=N_\pm$, which means that the two chiralities can be identified with the two types of fermions, with momenta close to either $+k_F$ or $-k_F$. The charges $Q_\pm$ then take, for both chiralities, integer values. In the interacting case, with $g\neq 1$, the situation is different. The distinction between the two types of chirality is no longer identical to the separation into two types of fermions, given by $\chi$, and in general the chiral parts of the fermion number no longer take integer values. Thus a fermion injected into the system at one of the fermi points will split into two parts, each carrying a fraction of the fermion number to the right and to the left, respectively. The interesting question is whether we should view this chiral separation of the two parts of the charge as a charge fractionalization effect or rather as a splitting of the probabilities for an integer charge to move either to the right or to the left. This means that the question is whether the fractional charge (for $g\neq 1$) that moves either to the right or to the left should be regarded as quantum mechanically sharp.

To discuss this question, it is useful to focus on a specific case, where a fermion is injected with momentum close to one of the Fermi point, \eg $k\approx +k_F$. According to the above discussion it should dynamically be separated in its two parts, which move in opposite directions away from where it has been injected. The charges of the two chiral components in this case are $Q_\pm=\half(1\pm g)$. The initial state can be viewed as formed by acting with the fermion creation operator on the ground state, and the fermion creation operator can further be decomposed in two operators that create each of the chiral components of the state. The chiral creation operators, when expressed in terms of the chiral fields, have the form \cite{Pham00}
\be{crea}
V_\pm(x)=\exp(i\pi Q_\pm\gQ_{\pm}(x))
\ee
where $V_+$ creates the right-moving component and $V_-$ the left-moving one. Furthermore, if we follow the reasoning of Ref.~\cite{Pham00}, the question of sharpness can be settled by examining the properties of these creation operators separately. Each of the operators creates, at the formal level, a chiral state with fractional charge, and by examining the algebra of the operators associated with each chirality the conclusion which is drawn in \cite{Pham00} is that the operators $V_\pm(x)$ create quasiparticles with sharply defined fractional charge which also satisfies fractional statistics.

However, there are reasons why we do not find this conclusion fully convincing, and that is a part of the motivation for our work \cite{LeinaasHorsdal09}, where  we have examined the question of charge fractionalization in a different way. Let me briefly discuss why we do not find the arguments based on the discussion of the properties of the operators $V_\pm(x)$ convincing.

Clearly the state that we consider can be expressed in terms of the operators $V_+$ and $V_-$, but only when these are used in combination. When used separately, on the other hand, the operators are not well-defined.  This is due to the periodic boundary conditions that the field operators have to satisfy, as a consequence of the fermionic character of the fundamental particles of the system. All well-defined operators on the Hilbert space of states have to satisfy these boundary conditions, and it is straight forward to check that the operators $V_\pm$ do not fulfill this obligation. In fact, since they are creation operators of {\em fractional} charge they necessarily do not preserve the spectrum of the charge operators $N$ and $J$, which is another way to see that they violate the boundary conditions.

Let me also point out that the charge operators $Q_\pm$ measure {\em global} charges, in the sense that they depend only on the integrated charges (\ie fermion numbers) $N$ and $J$. As a linear combination of these two operators they in fact necessarily are sharply defined for any state which is an eigenstate of $N$ and $J$, even if they take themselves fractional values. However, in the discussion of sharp fractional charges it is the {\em local} charges that are relevant rather than the global ones. Whereas the global charges, which correspond to the $q=0$ components of the charge densities, are restricted by the boundary conditions of the fields, the local charges, which are determined by the $q\neq 0$ components, are not subject to these restrictions. Also note that the $q=0$ components are not associated with any particular direction of motion and the division of the total charge into the chiral components therefore is somewhat arbitrary. This is different for the $q\neq 0$ components which are associated with a direction of motion.

I will then proceed to define local charge operators for each chirality. These can be extracted from the total charge density, which for $q\neq 0$ can be expressed in terms of the bosonic operators $b_q$ and $b_q^\dag$ in the following way,
\be{rhoq}
\gr_{ q}
&=&\sqrt{\frac{L|q|g}{2\pi }}( \,b_q+ \,b_{-q}^\dag)
\ee
This expression contains also information about the chiral components of the charge density, since the time evolution is determined by the fact that $b_q$ is a positive frequency operator and $b^\dag$ is a negative frequency operator. With the frequency given by $\gw=u|q|$ the space and time dependent charge density is
\be{chardens}
\gr(x,t)=\sum_q \sqrt{\frac{L|q|g}{2\pi }}( \,b_q e^{i(qx-u|q|t)}+ \,b_{-q}^\dag e^{i(qx+u|q|t)})
\ee
This gives for the positive and negative chirality components
\be{chircomp}
\gr^+_q =
\left\{
\begin{matrix}
 \sqrt{L|q|g/2\pi }\,b_q\quad q>0
\cr\cr
\sqrt{L|q|g/2\pi }\,b_{-q}^\dag \quad q<0
\end{matrix}
\right.
\quad\gr^-_q =
\left\{
\begin{matrix}
 \sqrt{L|q|g/2\pi }\,b_{-q}^\dag\quad q>0
\cr\cr
\sqrt{L|q|g/2\pi }\,b_{q} \quad q<0
\end{matrix}
\right.
\ee

At this point it may be useful to introduce a new variable $\gG$ to characterize the two types of chirality, with $\gG=+$ for the right-moving and $\gG=-$ for left-moving modes. We note that in the non-interacting case ($g=1$) $\gG$ is identical to the quantum number $\chi$ which identifies the two types of fermions, while in the interacting case ($g\neq 1$) they are different. By use of this new variable, the above expressions for the chiral charge densities can be compactified
\be{chircomp2}
 \gr^\gG_q = \sqrt{\frac{L|q|g}{2\pi }}\,\left[\gQ(\gG q)\,b_q+\gQ(-\gG q)\,b_{-q}^\dag\right]
\ee
with $\gQ(q)$ as the Heaviside step function. 

By use of the sampling function $f(x;a,b)$, introduced earlier in Eq.\pref{sampling}, the local chiral charge operators can be defined which measure charges in the interval $(-a/2,a/2)$,
\be{loc}
Q_\gG(a,b)=\int dx f(x;a,b) \gr^\gG(x)
={2\over L}\sum_{q>0} \sqrt{\frac{Lqg}{2\pi }}\,
\frac{\sin(aq/2)}{q}
e^{-{1\over 4}q^2b^2} (b_{\gG q}+b_{\gG q}^\dag)
\ee
We are interested in the expectation values and fluctuations of these for the state where an electron is injected with momentum close to one of the fermi points $\chi k_F$. To be more precise let us assume that the initial state after the injection has the form
\be{init}
\ket{\psi}= \int dx\, \phi(x)\, \psi_\chi^\dag(x)\,\ket G\equiv \Psi_\chi^\dag \ket G
\ee
 $\psi_\chi^\dag(x)$ as the fermion field for species $\chi$, $\phi(x)$ is the wave function of the injected function, and $\ket G$ is the ground state of the interacting many-fermion system. $\Psi_\chi^\dag$ is then the creation operator of the initial state which is characterized by fermion numbers $N_{\chi'}=\gd_{\chi \chi'}$. The expectation values of the chiral charge operators have the form
 \be{locmean}
\mean{Q_\gG(a,b)}={2\over L}\sum_{q>0} \sqrt{\frac{Lqg}{2\pi }}\,
\frac{\sin(aq/2)}{q}
e^{-{1\over 4}q^2b^2} \bra G \Psi_\chi(b_{\gG q}+b_{\gG q}^\dag)\Psi_\chi^\dag \ket G
\ee

The expectation value on the right-hand side of the equation is in fact straight forward to evaluate. This is because the ground state is annihilated by the operators $b_q$ and because the commutators between these operators and the fermion creation and annihilation operators also have a simple form, as a consequence of the fact that $b_q$ and $b_q^\dag$ are linear in the charge densities. 
I will here skip the details and refer to \cite{LeinaasHorsdal09} for these. Let me instead focus on the $q$ dependence of the sum. We may be interested in choosing $a$ sufficiently large so that the whole contribution from the fermion wave function is captured by the integral. That can be done in the following way. We first take the infinite length limit of the circle, $L\to\infty$, which makes the sum over $q$ into an integral, and next take the limit $a\to \infty$. In this limit the function $\sin(aq/2)$ essentially suppresses the contribution of the integral to arbitrary small values of $q$. The corresponding expectation value for the chiral charges, denoted by $\bar Q_\gG$, can then be written as
\be{charlim}
\bar Q_\gG&=&\lim_{a\to \infty}\left[{1\over \pi}\int_0^\infty dq \frac{\sin(aq/2)}{q}e^{-{1\over 4}q^2b^2}\right] 
\lim_{q\to 0^\gG} \left[\sqrt{\frac{Lqg}{2\pi }}
\bra G \Psi_\chi(b_{ q}+b_{ q}^\dag)\Psi_\chi^\dag \ket G\right]\nn
&=& \lim_{q\to 0^\gG} \left[\half\sqrt{\frac{Lqg}{2\pi }}
\bra G \Psi_\chi(b_{ q}+b_{ q}^\dag)\Psi_\chi^\dag \ket G\right]
\ee
By comparing with the earlier expressions for the chiral charge densities, this expression can in fact be recognized as the following limit of the expectation value of the charge densities
\be{qbar}
\bar Q_\gG= \lim_{q\to 0}\mean{\half(\gr^\gG_q+\gr_q^{\gG\dag})}
\ee
This expression should be compared to the mean value of the global charges, defined as $Q_\gG=\half(N+\gG J)$, which we may identify as the $q=0$ components of the charge densities
\be{glob}
\mean{Q_\gG}=\mean{\gr^\gG_{q=0}}
\ee
So the distinction between the total local charge defined by \pref{qbar} and the global charge defined by \pref{glob} is the same as the distinction between the $q=0$ component of the charge density and the $q\to 0$ limit \cite{HeinonenKohn87}.

In the case we consider here, with a fermion injected at one of the Fermi points, a further evaluation of the local charges shows that the two expectation values in fact are equal. For an initial state with $\chi=+$ we then have
\be{equal}
\bar Q_\gG=\mean {Q_\gG}=\half(1+\gG g)
\ee
This result, which means that the transition $q\to 0$ is continuous,  is easy to understand when we consider situations where the transition is {\em not} continuous. That happens when the initial state contains a constant background charge in addition to that of the ground state. Such a constant charge distribution will only affect the $q=0$ component and not the $q\neq 0$ components of the charge density. In the case discussed here the fermion charge is injected locally and no constant background is created. I will discuss also another case, where a state with a purely right-going component is created. In such a case there has to be a compensating background charge and therefore a discontinuity in the transition $q\to 0$ for the charge density. 

\section{The two-dimensional representation}
In order to make clearer the distinction between the two variables $\chi$ and $\gG$ and to illustrate the dynamical separation of the two chiral components of the initial state, it is useful to introduce a  two-dimensional representation of the many-fermion system. 
The charge densities in the one- and two-dimensional representations are related by the mapping
\be{twodim}
\gr_2(x,y)={\cal N}\int dx'\int d\xi e^{-{1\over l^2} [(x-x')^2+{1\over 4}\xi^2-iy\xi]}\gr_1(x'+{\xi\over 2}, x'-{\xi\over 2})
\ee  
with $\cal N$ as a normalization factor,  and $l$ as an unspecified constant with the dimension of length. The equation relates the {\em off-diagonal} matrix elements of the one-dimensional  charge density $\gr_1$ to the {\em diagonal} matrix elements of the two-dimensional density  $\gr_2$. The two-dimensional representation we may view as providing a phase-space description of the one-dimensional system, with the $y$ coordinate corresponding to the rescaled momentum $kl^2$.

As discussed in Ref.~\cite{Horsdal07} the two-dimensional description can also be viewed as corresponding to an alternative, physical realization of the theory. It describes a  two-dimensional electron gas in a strong magnetic field, with $l$ as the magnetic length.   With the electrons restricted to the lowest Landau level, the electron gas is dynamically equivalent to one-dimensional system, and with a harmonic confinement potential in one direction of the plane,  the Hamiltonian is equivalent to that of the one-dimensional fermion system with a quadratic dispersion. (The same correspondence has recently been applied in \cite{Berg09}, where the possibility of detecting charge fractionalization on the edge of a quantum Hall system has been analyzed.)  

\begin{figure}[h]
\begin{center}
\includegraphics[width=15cm]{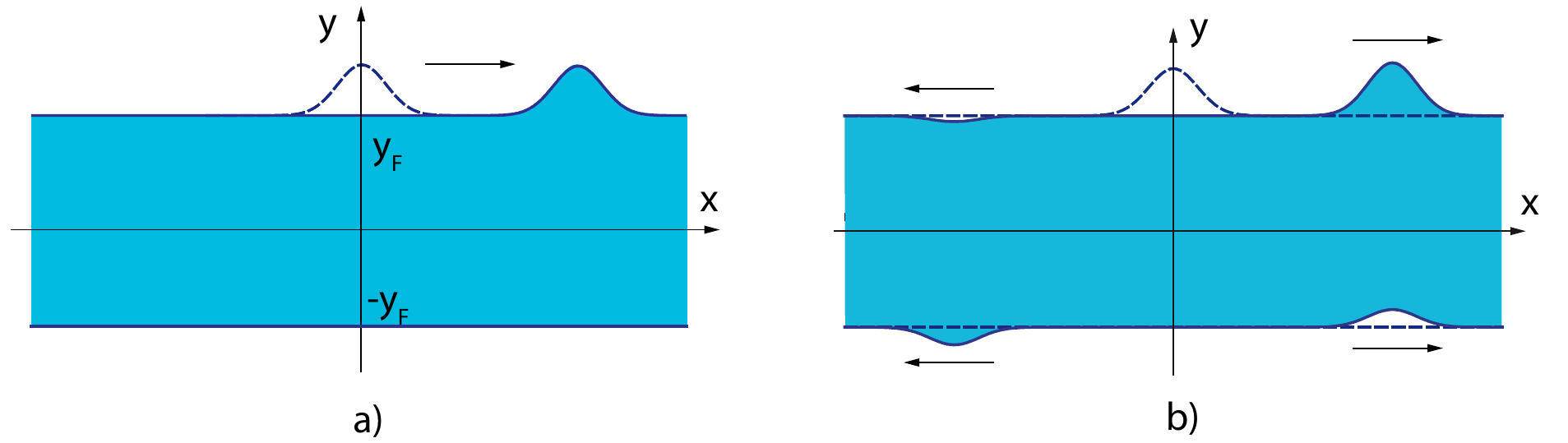}
\end{center}
\caption{\small Two-dimensional representation of the fermion system, with the $y$-coordinate proportional to the momentum $k$. The ground state corresponds to a state with constant density in a band between the Fermi momenta, here represented by $\pm y_F$. If a Fermion is injected in the non-interacting system ($g=1$) at the upper edge (dashed curve), the excitation will propagate as a modulation of the edge to the right, as illustrated in the left part a) of the figure. If there is an inter-edge interaction ($g\neq 1$), a fermion injected on the upper edge will instead separate in a right- and a left-moving mode, and each of these will have components on both edges, as illustrated in the right part b) of the figure.  \label{KiralSep1}}
\end{figure}

The ground state of the non-interacting system corresponds to the situation where a band of electron states along the $x$-axis are fully occupied, with sharp transition to the unoccupied states along the edges of the band. Interaction between the electrons introduces a modification of the edge profile, but does not change the density of the interior away from full filling (of the lowest Landau level). The low energy approximation, in this picture, corresponds to a description of the dynamics restricted to edge excitations of the system. When the electron interaction is sufficiently long range to interconnect the two edges of the system, we have $g\neq 1$ for the interaction parameter, and this corresponds to the situation with a coupling between the edge excitations on the two sides of the band. 

In the two-dimensional representation the parameter $\gG$ can be associated with the horizontal ($x$)-direction, where $\gG=+$ corresponds to propagation to the right and $\gG=-$ corresponds to propagation to the left.  The parameter $\chi$, on the other hand, is associated with the vertical ($y$) direction, with $\chi=+$ corresponding to the upper edge and $\chi=-$ corresponding to excitations on the lower edge. 

The situation is illustrated in Fig.\ref{KiralSep1}, where an electron is injected close to $x=0$ on one of the edges, pictured in the form of a modulation of the edge. In the case $g=1$ (Fig.\ref{KiralSep1}a) with no coupling between the edges, we have $\chi=\pm$ equivalent to $\gG=\pm$, and the modulation is therefore displaced with time in the positive $x$-direction for injection on the upper edge and in the negative direction for injection on the lower edge. In the interacting case, with $g\neq 1$, the situation is different due to the coupling between the two edges. The charge moving to the right on the upper edge will then produce a (smaller) image charge moving to the right on the lower edge (Fig.\ref{KiralSep1}b). The ratio between these two charges is uniquely determined by the interaction parameter $g$. However, charge conservation on the lower edge implies that there must be a compensating charge of equal size but opposite sign that moves to the left. This in turn has a mirror charge sitting on the upper edge that also moves to the left. 

As indicated in the figure the initial charge is in this way separated in four parts. Let us denote these charges by $Q_{\chi\gG}$, so that $N_\chi=\sum_\gG Q_{\chi\gG}$ and $Q_\gG=\sum_\chi Q_{\chi\gG}$. It is interesting to note that these four components of the injected charge are all determined by charge conservation on each edge combined with the given ratio between the right (left)-going charge on the upper (lower) edge and its mirror component on the opposite edge.  Thus, the ratios between the primary charge, defined by $\chi=\gG$ and its mirror charge, defined by $\chi=-\gG$ is given by 
\be{ratio}
\frac{Q_{++}}{Q_{-+}}=\frac{Q_{--}}{Q_{+-}}={{g+1}\over{g-1}}
\ee
This fixed relation for the {\em local} mirror charges on the two edges can  be ascribed to the fact that $g$ acts as a mixing parameter between the two original chiralities $\chi=\pm$ when the interaction between the edges is turned on. With the given initial conditions $N_+=1$ and $N_-=0$ we in addition have
\be{incon}
Q_{++}+Q_{+-}=1\,,\quad Q_{-+}+Q_{--}=0
\ee
These two sets of conditions give
\be{results}
 &&Q_{++}={1\over{4g}}(g+1)^2\,,\quad \;\;\;Q_{-+}={1\over{4g}}(g^2-1)\nn
 &&Q_{--}=-{1\over{4g}}(g^2-1)\,,\quad  Q_{+-}=-{1\over{4g}}(g-1)^2
\ee
which for the two chiral charge components implies
\be{chcharge}
Q_+=Q_{++}+Q_{-+}=\half(1+g)\nn
Q_-=Q_{+-}+Q_{--}=\half(1-g)
\ee
This is consistent with the results of the more detailed calculation sketched in the previous section.

As already pointed out, in the case with a sudden injection of a fermion on one of the edges, there is no difference between the expectation values of the local and global charges. There is, however, a difference when we instead consider the quantum fluctuations of these variables. I will soon discuss this point further, but let me first discuss how to choose initial conditions so that only one chiral component with fractional charge will be present. I have argued that this cannot be done by introducing creation operators of the form \pref{crea}, since these do not treat the background charge correctly. The point to note is that a single local excitation with fractional charge has always be accompanied by a compensating background charge.

A constructive way to introduce such an initial state is the following. We consider a fermion injected locally in the {\em non-interacting} system at one of the edges. Since $g=1$ this excitation has well-defined chirality. As the next step we slowly (adiabatically) turn on the interaction so that $g\neq 1$. If this change is sufficiently slow it will not create a local excitation with opposite chirality ( a back-scattered component) but will instead create a compensating background charge uniformly distributed over the system. 
\begin{figure}[h]
\begin{center}
\includegraphics[width=8cm]{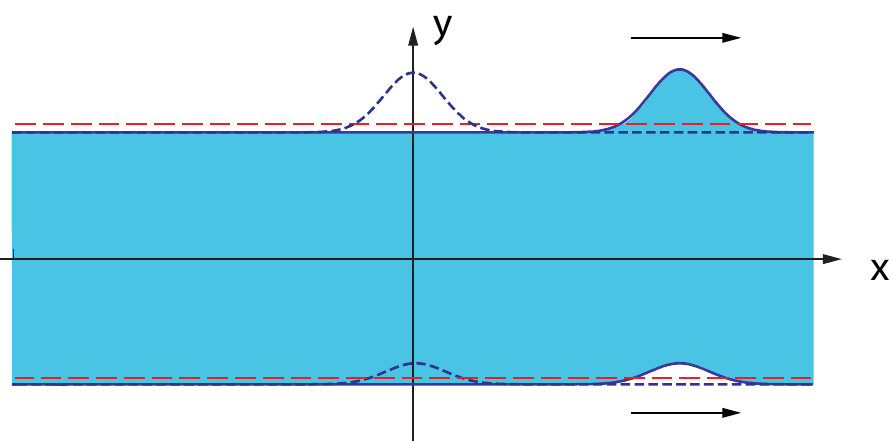}
\end{center}
\caption{\small  Creation of a purely right-moving mode, shown in the two-dimensional representation. A fermion is injected into the non-interacting system on the upper edge and the interaction is adiabatically turned on. The created modulation of the edges (blue dashed curves), which carries a fractional charge, will propagate to the right. A compensating, constant background charge is created when the interaction is turned on, here illustrated by the difference between the original positions of the edges (red dashed lines) and the positions of the edges with the excitation (solid lines).\label{KiralSep2}}
\end{figure}
The initial state can now be written
\be{ad}
\ket{\psi} = U_g\Psi_\chi^\dag \ket F
\ee
with $\ket F$ as the ground state of the non-interacting system and $U_g$ as the operator that performs the adiabatic turning on of the inter-edge interaction. The evaluation of the two chiral components of the charge can be done in much the same way as the first case. This is so since $U_g$ can be identified with the unitary operator that transforms the bosonic creation and annihilation operators of the non-interacting system into the corresponding operators of the interacting system \cite{LeinaasHorsdal09}.  I only quote the result for the expectation values of the local charges 
\be{meanval}
\bar Q_+= \sqrt{g}\,,\quad \bar Q_-= 0
\ee
which shows that for a particle injected on the upper edge only a right-moving excitation is created. These expressions is to be compared with the corresponding ones for the {\em global} charges $Q_\pm$ defined in \pref{zerofields}. They take the same values as before, since they are fixed by the fermion numbers $N_+=1, N_-=0$, which are not affected by the interaction,
\be{global}
Q_+= \half(1+g)\,,\quad Q_-= \half(1-g)
\ee
There is now a difference between the local charges \pref{meanval} and the global charges \pref{global} which can be ascribed to the presence of constant background charges on each edge. These have been created by the adiabatic process, under which the original integer charges on the two edges are separated in the fractional, local components and the compensating background charges. The situation is pictured in Fig.\ref{KiralSep2}. 

It is interesting to note that the non-integer charge carried by the positive chirality component in the adiabatic case is different from the charge of the same chirality in the case of sudden injection. This shows explicitly that even if non-integer charges may be created in the Luttinger liquid, the value of these charges are not uniquely determined by the theory, but rather depends on the initial conditions under which the fractionally charged excitations have been created.

\section{Charge fluctuations}
Let me finally turn to the question of sharpness of the fractional charges that are created in the Luttinger liquid. I will then focus on the two examples discussed above, with sudden and with adiabatic injection of a fermion charge at one of the edges. Since the system is gapless we expect, in the same way as already discussed, that there are ground state fluctuations that cannot be suppressed by a smooth sampling function. This point can, in the bosonized representation of the theory, be checked for an arbitrary value of the interaction parameter $g$. The quantity to be evaluated is the variance
\be{fluct}
\gD Q(a,b)^2=\bra G  Q(a,b)^2 \ket G
\ee
and this can be done in essentially the same way as sketched for the determination of the expectation values of the fractional charges. Again I refer to \cite{LeinaasHorsdal09} for details, and simply quote the result 
\be{gsfluct2}
\gD  Q_{\pm}(a,b)^2={ g\over 4\pi^2} {a^2\over b^2}\;_2F_2 (1, 1; 3/2, 2; - a^2 / 2b^2)\equiv \half \gD  Q_{0}(a,b)^2 
\ee
where $\gD  Q_{0}(a,b)^2$ is the variance of the full local ground state charge. We note that this variance is, except for the presence of $g$ as a scaling parameter, precisely the same as previously found for the massless Dirac vacuum in one dimension. That is in fact not so surprising, since the linearized theory underlying the bosonized description can be viewed as describing a system of massless Dirac particles. For $a>>b$ the variance then has a logarithmic dependence on the ratio $a/b$, as earlier discussed.

The presence of undamped ground state fluctuations implies that {\em any} excited state of the system will also show such charge fluctuations for large values of $a/b$.  This implies that in the gapless system that we consider a local charge cannot be quantum mechanically sharp in the same sense as in the gapped system.  However, it i possible to measure the fluctuations {\em relative} to the ground state fluctuations in the following way. We write the variance of the charge operator for a general state as
\be{gsfluct3}
\gD  Q_{\Gamma}(a,b)^2=\half \gD  Q_{0}(a,b)^2+\Lambda_\gG(a,b)
\ee
where $\Lambda_\gG(a,b)$ measures the deviations from the ground state fluctuations. This deviation remains finite when $a/b\to\infty$ for any state which carries a finite excitation charge. Following \cite{LeinaasHorsdal09} we may then {\em define} a chiral charge to be sharp when this deviation from the background variance vanishes for sufficiently large $a$.

The results we have found, for both cases, with either sudden injection of a fermion charge or with adiabatic switching on the interaction, is that the deviations $\Lambda_\gG(a,b)$ vanish in the limit $a\to \infty$. Thus, in both cases the fractional charges created dynamically by the chiral separation are sharp in the meaning given above. This is so even if the expectation values of the chiral charges in the two cases are different.

\section{Concluding remarks}
The idea of charge fractionalization in Luttinger liquids is an interesting one, but as discussed in this talk the appearance of non-integer charges due to chiral separation is in some important respects different from charge fractionalization in the more ``classic" sense. The main reason for this is that Luttinger liquids are gapless systems, where charge fluctuations of the many-particle background cannot be controlled by a smooth sampling function for the local charge operator, in the same way as in gapped systems.  However, in the examples considered here, with a fermion injected close to one of the Fermi points, the fractional charges are sharp in the sense that the fluctuations are identical to the background fluctuations of the many-particle ground state. This is in contrast to what happens with more general initial states. In particular, if a particle is injected with a non-vanishing probability for exciting both edges, then there will be a finite addition to the variance of the chiral charges, which may be interpreted as statistical mixture of two different distributions of the charge.

As I have stressed, it is important in the discussion of sharpness of the charges to focus on {\em local} charge operators rather than the global ones. For a fractional charge the difference between these is a compensating charge,  which in a compact space will be evenly distributed over the background, while it in an open system may be collected at the remote edges of the system. For the charge density this means that the information about the fractional charge, and in particular the quantum fluctuations of the charge, sits in the Fourier components $q\neq 0$ rather than in the component $q=0$. This distinction between the two types of contributions is also important since the $q=0$ component is constrained by the conserved fermion numbers, and is insensitive to the inter-edge interactions, while the $q\neq 0$ components are not restricted in the same way but are sensitive to the interactions.  

The point that fractional charges with different values can be created by choosing different initial conditions is also making the fractionalization effect in the Luttinger liquid different from the effect in the gapped systems. In those cases the topological properties of the systems impose restrictions on the allowed charge values, but in the Luttinger liquid there seems to be no similar restriction. The boundary conditions of the fields can be viewed as a topological constraint, but these are non-dynamical, fixed to the fundamental fermion charges, and do not impose constraints on the fractional, local charges. In fact, even in the non-interacting case $g=1$ fractional charges of {\em arbitrary} values can be created as polarization charges induced by an external local potential \cite{LeinaasHorsdal09}. If such a charge is attracted to a small region by the potential and then released by suddenly turning off the potential, then the charge will separate in the two chiral components, which will generally carry  fractional charges with values determined by the strength of the potential. And these fractional charges will be quantum mechanically sharp in the same sense as in the two other examples \cite{LeinaasHorsdal09}. In this case (with $g=1$) the dynamical separation of the two different chiralities is {\em not} the cause of the fractionalization.

These results, that fractional charges of arbitrary value can be created that are sharp in the meaning given above, makes it difficult to identify these as charges of elementary quasiparticle excitations of the system. A possible conclusion to draw may then be that charge fractionalization in Luttinger liquids is a different, and simply a less interesting phenomenon than in gapped systems, where the fractionalization is associated with the appearance of quasiparticles with exotic properties. However, maybe one should be open for more interesting possibilities.
Let us compare the situation for the Luttinger liquid with that of gapless systems described by Fermi liquid theory, in dimensions higher than one. In these systems the quasiparticle picture is of fundamental importance, even if the idea of charge fractionalization seems not to be a central issue. However, as discussed in Ref.\cite{HeinonenKohn87}, also in Fermi liquids the quasiparticles may be considered as carrying non-integer charges. Thus, if these are considered to evolve from the fundamental particles (electrons) through an adiabatic process, they seem to carry only a fraction of the electron charge. This can be shown in a perturbative analysis, where  a compensating charge is stored in the background, in much the same way as discussed here \cite{HeinonenKohn87}. For the Fermi liquids such a charge fractionalization seems, however, less important than the fact that the quasiparticles appear as weakly interacting particles that satisfy Fermi-Dirac statistics like the electrons.

As a final remark let me therefore suggest that there may be some additional conditions that are missing in the discussion of fractional charges in Luttinger liquids, which could be of importance for the application of the quasiparticle picture to this system. These conditions could imply that there are quasiparticles that carry specific, non-integer charges, but more importantly these quasiparticles could be different from the fundamental particles in the sense that they are weakly interacting and obey {\em fractional statistics} rather than the Fermi-Dirac statistics. This would bring the conclusion closer to ideas discussed in Refs.\cite{Isakov98,Wu00} about relations between Luttinger liquid theory and a free gas of exclusion statistics particles.\\

\noi
{\bf Acknowledgement}\\
Thanks to Mats Horsdal and Hans Hansson for useful comments on the manuscript.

 
\end{document}